\documentclass[preprint,showpacs,nofootinbib,prd,aps]{revtex4-1}
\usepackage[utf8]{inputenc}
\usepackage{graphicx,amstext,amssymb}
\pdfoutput=1 

\begin{document}
\title{Modified  Skellam,  Poisson and Gaussian distributions in semi-open systems at charge-like conservation law}

\author{Yu.~M.~Sinyukov}
\affiliation{Bogolyubov Institute for Theoretical Physics,
03680 Kiev,  Ukraine}

\begin{abstract}

A modification of the Skellam and Poisson distributions is proposed for subsystems 
when the constraints imposed by the  charge conservation law in the complete system are taken into account. Such
distributions can be applied, for example, for an analysis of the fluctuations of baryon and net baryon
numbers in certain pseudo-rapidity interval in $A+A$ and $p+p$ collisions with high multiplicities.
The presented modified Skellam, Poisson and Gaussian distributions can be utilized also in various
branches of science, when one studies the  fluctuations of the two variables related to a subsystem, as well as the distribution of the difference of these variables, while the mentioned difference in the total system is fixed.
          
\end{abstract}


 \maketitle

 \section{Introduction}

 The famous Poisson distribution in statistics appears in 1837 \cite{Po}. More than one century later  J. G. Skellam published the paper  "The frequency distribution of the difference between two Poisson variates belonging to different populations" \cite{Sk}. Both the Poisson and Skellam distributions are widely utilized for an analysis of very different phenomena, in various branches of sciences. Despite our approach is quite general, to be concrete, this note we address to the actual and relatively new  field of science - relativistic nucleus-nucleus and proton-proton collisions with high  multiplicity at, particularly, CERN LHC and BNL RHIC. In these experiments an extremely high  energy of colliding nucleons converts into  multiparticle and  multicomponent systems. The successful analysis of the particle production for different species within the  statistical models \cite{pbm1,Becat,Clemans}, where a minimal number of parameters, such as the temperature and chemical potentials are almost enough to describe and predict the data, indicates a thermal nature of the final state formed in multiple  elementary processes accompanied the collisions. Therefore the methods of statistical physics, that utilized micro-canonical, canonical and grand canonical ensembles to described the formed systems are widely used.

Typically,  only a part of the system is acceptable for the direct data analysis at current LHC or RHIC detectors. It is clear, that in the  case of full acceptance, the so-called 4$\pi$ geometry, the {\it net} baryon number in the final state that is equal to initial one -  number of nucleons  in the two colliding nuclei or just value 2 in $p+p$ collisions -- is fixed and not fluctuated. In small subsystems of the total final system, the distributions of baryons and anti-baryons can be both approximated by the Poissonian ones, then difference between the corresponding particle numbers (the net baryon number) are distributed according to related Skellam  function \cite{Sk}. It was checked by the ALICE Collaboration \cite{Anar} that such a situation really takes place in relatively small pseudo-rapidity interval in $Pb+Pb$ collisions. In the case, when subsystem of baryons + anti-baryons, that is available for analysis, is  not small and comparable to the total system - such a situation takes place in nucleus-nucleus collisions with relatively low energies (BES at RHIC, FAIR and NICA planning experiments) - the baryon charge conservation law will certainly deform both the Poisson and Skellam distributions for  baryon/anti-baryon numbers and net baryon charge in the subsystem. 
 
Despite the fact that in some publications \cite{goren, pbm, koch, asak}, the  baryon number conservation law in such {\it  semi-open} systems is discussed to some extend, the problem as whole is not solved up to now. In this note we start from generalization of the Skellam distribution; the corresponding modification of the  Poisson one will follow from the former in straightforward way.

\section{The statement of the problem }

Suppose that independent discrete values $n_i$ are distributed according to Poisson,  $P(N_i,n_i)$, with the mean values $N_i$:  
\begin{eqnarray}
P(N_i,n_i)= \frac{N_i^{n_i}}{n_i!}\exp(-N_i) \label{Poisson}
\end{eqnarray}
If there are several uncorrelated subsystems, then  the sum of discrete values, $\sum{ n_i}$, can be presented again as the Poisson distribution $P(\sum{N_i},\sum {n_i})$.   Let us consider a system, consisting of  particles of two species: say, baryons $(N,n)$ and anti-barions (${\overline N},{\overline n}$). Since the distributions are of the Poissonian type and mutually independent, one can calculate distribution $p(k)$ of difference $k$ between particle numbers in both components, in our case,  baryon minus antibaryon numbers $k=n-{\overline n}$: $p(k;N,{\overline N})= \sum_{{\overline n}}^{\infty}P(N,{\overline n}+k)P({\overline N},{\overline n})$. The latter is expressed by the Skellam distribution \cite{Sk}:
\begin{eqnarray}
p(k; N,{\overline N}) = \exp(-N-{\overline N})\left(\frac{N}{{\overline N}}\right)^{k/2} I_k(2\sqrt{N{\overline N}}) \label{Skellam}
\end{eqnarray}
Here $I_k$ is the modified Bessel  function of the first kind.
The distribution is normalized:
\begin{eqnarray}
\sum_{k=-\infty}^{\infty} p(k; N,{\overline N}) =1 \label{norm}
\end{eqnarray}
The mean value $m$ is
\begin{eqnarray}
m=\sum_{k=-\infty}^{\infty} kp(k; N,{\overline N}) =N-{\overline N}. \label{m}
\end{eqnarray}

\begin{figure}
 \centering
  \begin{minipage}[c]{.80\linewidth}
		\centering
     \caption{The total system. Any selected sybsystem $i$, say ``1'', \\is complemented by system ``2'' (and wise-verse) to a complete \\closed system.}
\label{fig:ill1}
\end{minipage}%
\begin{minipage}[c]{.20\linewidth}
     \centering
      \includegraphics[width=\linewidth]{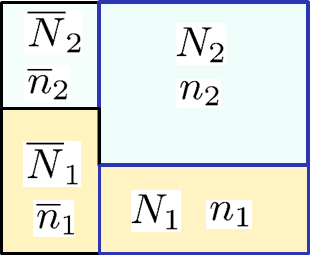}
\end{minipage}
\end{figure}

Let us input constraint for this baryon-antibaryon system and associate it with the baryon number conservation in the isolated complete system. Namely, we consider the ensemble of the systems under the constraint that all the systems in the ensemble have exactly the same net baryon number $B$, for other observables the averaged values are fixed only. Such an ensemble can be realized, for instance,  in very central nucleus-nucleus collisions. Then the total final multiplicity can fluctuate, but fluctuations of the net baryon number in the total system are completely suppressed.  Let us divide the total system into the two $i$-subsystems, $i=1,2$, with the mean numbers \{$N_1,{\overline N}_1$\} and \{$N_2,{\overline N}_2$\}, Fig. \ref{fig:ill1}.  
The fluctuations of particle numbers, baryon $n_i$ and anti-baryon $\bar{n}_i$, in the two subsystems  are not independent anymore because the conservation law constraint, $n-{\overline n}=N-{\overline N} = B = const$ ~ ($n=\sum_i n_i,~{\overline n}=\sum_i {\overline n}_i$),  
brings  the condition for the two components \{$n_1,{\overline n}_1$\} and \{$n_2,{\overline n}_2$\}, as well as for the differences $k_1=n_1-{\overline n}_1$ and $k_2=n_2-{\overline n}_2$, in both subsystems:
\begin{equation}
B=n_1+n_2-{\overline n}_1-{\overline n}_2 = k_1+k_2 = N_1 + N_2 - {\overline N}_1 - {\overline N}_2  
\label{B}
\end{equation}

Our aim is to generalize the Skellam distribution (\ref{Skellam}) for the semi-open $i$-systems, $i=1$ or 2. One of the possiblities to do this, is just to construct some  function ${\cal F}$ by multiplying  the product  of independent distributions  by the  Kronecker  $\delta$-function, e.g., ${\cal F}= P(N_1,n_1)P(N_2,n_2)P(\bar{N}_1,\bar{n}_1)P(\bar{N}_2,\bar{n}_2)\delta_{n-{\overline n}}^B$, or ${\cal F}=P(\sum{N_i},\sum {n_i})P(\bar{N}_1,\bar{n}_1)P(\bar{N}_2,\bar{n}_2)\delta_{n-{\overline n}}^B$, or
${\cal F}=P(N_1,n_1)P(N_2,n_2)P(\sum \bar{N}_i,\sum \bar{n}_i)\delta_{n-{\overline n}}^B$, or half-sum of the last two expressions, etc. \footnote{e.g., in Ref. \cite{goren} the Kronecker-delta constraint is imposed for the product of the partition functions of initially  independent subsystems.}. After that a Skellam-like distrubution is presented for subsystem $i$ as
\begin{equation}
p(k_i; N_i,{\overline N}_i)=\sum_{n_j,{\overline n}_l }{\cal F}(\{n_j, \bar{n}_l\})\delta_{n_i-{\overline n}_i}^{k_i}
\label{F}
\end{equation}
The resulting distribution  will  depend on the choice of the initial  form ${\cal F}$. It cannot be derived unambiguously without a  microscopic model/theory that includes mechanisms ensuring the charge conservation law in the total system.  

    Our goal is to find the simplest analytic approximation to this problem which is based on general constraints required for such semi-open systems. We will compare our results with the model based on the binomial distribution (see \cite{pbm}, \cite{koch}) with probabilities $q$ and  ${\overline q}$  for baryon and anti-baryon  to belong to certain subsystem, say ``1'', when the total numbers $n$ and ${\overline n}$ in complete system are given. Let now these numbers $n$ and ${\overline n}$  fluctuate  with the Poisson distributions having the mean values $D$ and ${\overline D}$ correspondingly. In such a model ${\cal F}$ in Eq. (\ref{F}) is
\begin{eqnarray}
{\cal F}(\{n_j, \bar{n}_l\})=C\sum_{n,{\overline n}}\delta_{n-{\overline n}}^B \delta_{n_1+n_2}^n\delta_{{\overline n}_1+{\overline n}_2}^{{\overline n}}   P(D,n)P({\overline D},{\overline n}) \nonumber \\ \times ~  q^{n_1}(1-q)^{n_2}{\overline q}^{{\overline n_1}}(1-{\overline q})^{{\overline n_2}}\frac{n!}{n_1!~n_2!}\frac{{\overline n}!}{{\overline n_1}!~{\overline n_2}!}
\label{bin}
\end{eqnarray}
where $C$ is the normalization constant and
\begin{equation} 
q=\left\langle n_1 \right\rangle/\left\langle n \right\rangle =N_1/N,~{\overline q}=\left\langle {\overline n_1} \right\rangle/\left\langle {\overline n} \right\rangle ={\overline N_1}/{\overline N}.
\label{q}
\end{equation} Only the numerical calculations are possible to  provide summation of infinite series of hypergeometric functions in order to find from (\ref{bin}) the distribution over $k_1=n_1-{\overline n}_1$ according to Eq. (\ref{F}). 
Suddenly imposed constraint $\delta_{n-{\overline n}}^B$ dramatically changes the initial Poissonian-based ``bare'' distributions for $n_j$ or $\bar{n}_l$. Correspondingly, the mean values $N =\left\langle n\right\rangle$, ${\overline N} =\left\langle{\overline n}
\right\rangle$ are differ from the ``bare'' mean values ($D, {\overline D}$ in Eq. (\ref{bin})), and their connection is expressed through the  sum of hypergeometric functions. Nevertheless, to compare our results with results based, for certainty, on the binomial-Poissonian distribution (\ref{bin})  we provide such calculations.

\section{Generalization of the Skellam distribution}

In this note we propose the natural generalization of the Skellam and Poisson distributions that not deal with the ``bare'' distributions and corresponding ``bare'' mean values but are expressed analytically directly through experimentally observed particle numbers and its mean values for the system and subsystems. Pursuing this aim let us present the modified Skellam distribution for subsystem $i$ in the following generalized form:
\begin{equation}
\widetilde{p}(k_i; N_i,{\overline N}_i) = e^{-(M_i+{\overline M_i})}\left(\frac{M_i}{{\overline M_i}}\right)^{(k_i-k_i^0)/2}
I_{k_i-k_i^0}(2\sqrt{M_i{\overline M}_i}~) 
\label{modSkellam}
\end{equation}
To fix the expressions for $M_i, {\overline M}_i, k_i^0$ in the ansatz (\ref{modSkellam}) let us list the conditions for the modified Skellam distribution:
\begin{enumerate}
\item The normalization condition (\ref{norm}) with substitutions $p\rightarrow {\tilde p}$, $k \rightarrow k_i$, and $\{N,{\overline N}\} \rightarrow \{N_i,{\overline N}_i\}$  must be satisfied for $\widetilde{p}(k_i; N_i,{\overline N}_i)$. 
\item The equality  (\ref{m}) for the mean value must be satisfied under the above substitutions.
\item Because of the symmetry of the conservation law constraint (\ref{B}) with respect to the  mutual permutations $N_1 \leftrightarrow N_2$,  ${\overline N}_1 \leftrightarrow {\overline N}_2$   and $k_1 \leftrightarrow k_2$, the analytical expressions for modified Skellam distributions for the two subsystems should transform to each other  $\widetilde{p}(k_1; N_1,{\overline N}_1) \leftrightarrow  \widetilde{p}(k_2; N_2,{\overline N}_2)$ under these permutations.
\item If one of the subsystems (say, ``1'') is much smaller than the other one, $N_1 + {\overline N}_1 \ll N_2,{\overline N}_2$, then fluctuations of the two components, baryon and anti-baryon, in the subsystem ``1'' are uncorrelated Poissonian ones (the second subsystem just plays the role of a ``thermal bath''). So the  distribution (\ref{modSkellam}) tends to the Skellam one (\ref{Skellam}), 
\begin{equation}
\widetilde{p}(k_1; N_1,{\overline N}_1) \rightarrow  p (k_1; N_1,{\overline N}_1),~  N_1 + {\overline N}_1 \ll N_2,{\overline N}_2,
\label{Skellam limit}
\end{equation}
see Fig. \ref{fig:ill2}.  The situation is wise-verse as for the permutation $``1'' \leftrightarrow ``2''$.
\begin{figure}
   \centering
    \begin{minipage}[c]{.80\linewidth}
     \centering
     \caption{Illustration to items (C4), (C5) and Eqs.(\ref{Skellam limit}), (\ref{delta}).}
\label{fig:ill2}
\end{minipage}%
\begin{minipage}[c]{.2\linewidth}
     \centering
      \includegraphics[width=\linewidth]{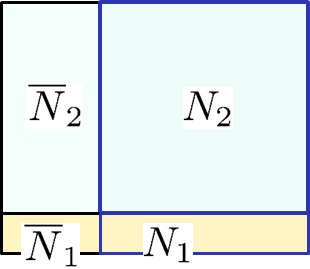}
\end{minipage}
\end{figure} 
\item When one of the subsystems (say, ``1'') vanishes,  $N_1,{\overline N}_1 \rightarrow 0$, the subsystem ``2'' occupies, in fact, the total system, $N_2-{\overline N}_2 = B$, and then, according to the net baryon charge conservation law,  
\begin{equation}
\widetilde{p}(k_2; N_2,{\overline N}_2) \rightarrow \delta^B_{k_2},~ N_2\rightarrow N,~{\overline N}_2 \rightarrow {\overline N}.
\label{delta}
\end{equation}
The situation must be, of course,  wise-verse when one permutes the systems,  $``1'' \leftrightarrow ``2''$.

\item One more restricting condition appears if the total system has only one-component, e.g. when ${\overline N}_i\rightarrow 0$ for both $i=1,2$, see Fig. \ref{fig:ill3}.
 
\begin{figure}[!ht]
   \centering
    \begin{minipage}[c]{.50\linewidth}
     \centering
     \caption{Illustration to items (C6).}
\label{fig:ill3}
\end{minipage}%
\begin{minipage}[c]{.25\linewidth}
     \centering
      \includegraphics[width=\linewidth]{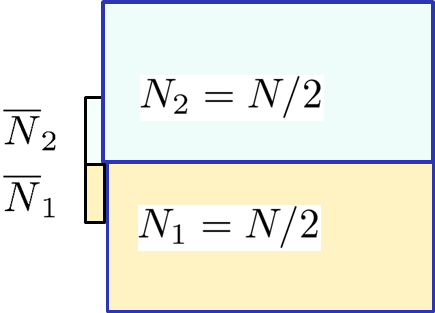}
\end{minipage}
\end{figure} 

Then the fluctuations in $k_i=n_i-{\overline n}_i=n_i$ inside the selected $i$-subsystem arise only because of the fluctuations of the {\it baryons} between the subsystems ``1'' and ``2''. It is obviously, that when $N_1=N_2=N/2$, the relative fluctuations, $\sigma_i/N_i$, in any single subsystem  will be twice suppressed as compare to the Skellam $\stackrel{{\overline N}_i=0}{\rightarrow}$ Poisson one in an independent subsystems. The  fluctuation in the single $i-$subsystem is extended to the entire system $N$: it enforces the same fluctuation (with opposite sign) in the other subsystem  because of the charge conservation law.  The dispersion of fluctuations related to (\ref{modSkellam})  is defined, similar as in the  Skellam case,  by $\sigma=\sqrt{M_1+M_2}$  (see below, Eq. ({\ref{sigma}})) and in {\it independent} Poisson subsystems,  when $M_i= N/2$, is $\sigma_{{ind}}=\sqrt{N}$. So, to get  $\sigma= \sigma_{ind}/2$ when the conservation law constraint is imposed, one should put $M_i=N/8$. On the other hand, if $N_i \rightarrow N$, it must be: ${\widetilde p} \rightarrow \delta_{k_i=n_i}^B$.

\end{enumerate} 

As we will find, these six obvious conditions are enough to define a simplest expression for  the modified Skellam distribution (\ref {modSkellam}). In what follows we will refer to these conditions as (C1), (C2), (C3), (C4), (C5), (C6).  

For the compactness of the subsequent presentation in an analysis of the items (C1)-(C5), we temporary rename the notations: $M_i,{\overline M}_i \rightarrow G_i,{\overline G}_i$.  The normalization condition (C1) are satisfied when $G_i,{\overline G}_i\geq 0$ and $k^0_i$ is integer. This can be seen immediately when one changes the  summation variable: $k_i\rightarrow q_i= k_i-k_i^0$. In what follows, using the concrete expressions for $k^0_i$, we always imply the nearest  integer numbers to the corresponding values. With the same changing of the summation variable, one can find that condition (C2) brings us the expression for $k_i^0$:
\begin{equation}
k_i^0=N_i + {\overline G}_i -{\overline N_i}-G_i  
\label{k0}
 \end{equation} 

  To guarantee condition (C3) -- the invariant form  of the modified distribution under permutation $``1''\leftrightarrow ``2''$  -- let us transform  $\widetilde{p}(k_1; N_1,{\overline N}_1)$ into $\widetilde{p}(k_2; N_2,{\overline N}_2)$ by means of Eqs. (\ref{B}) and (\ref{k0}). Then one has  $k_1-k_1^0 = -(k_2-k_2^0)+G_1-{\overline G}_1+G_2-{\overline G}_2$.  Putting $G_1+G_2={\overline G}_1+{\overline G}_2$ one gets $k_1-k_1^0 = -(k_2-k_2^0)$ and also transforms  the exponent $e^{-G_1-{\overline G}_1}$ into $e^{-G_2-{\overline G}_2}$ in (\ref{modSkellam}). Then, accounting for Bessel function property $I_n(z)=I_{-n}(z)$ and negative sign before the expression $(k_2-k_2^0)$ after transformation of $k_1-k_1^0$, one gets the final result to fulfill the condition (C3):
\begin{equation}
G_1={\overline G}_2, ~ G_2={\overline G}_1
\label{symM}
 \end{equation} 
  
 Now let us present $G_i,~{\overline G}_i$ in the form $G_i=\alpha_i N_i$ and ${\overline G}_i={\overline \alpha}_i {\overline N}_i$. The limits described by the condition (C4) in the situation  when, say subsystem $''N_1+{\overline N}_1~''$ is much smaller than both of components, $N_2$ and ${\overline N}_2$, of the subsystem $``2''$, require that $\alpha_1\rightarrow 1$ and ${\overline \alpha}_1\rightarrow 1$ in this case. Then the modified Skellam distribution tends to standard one according to Eq. (\ref{Skellam limit}). Similarly for the second system.
To satisfy the condition (C5) when system ``2'' tends to be the total system, and so $N_1\rightarrow 0$ and ${\overline N}_1 \rightarrow 0$, one must to put  $G_2\rightarrow 0$ and ${\overline G}_2 \rightarrow 0$. Then $k_2=k_2^0=N_2-{\overline N}_2$ and the Bessel function in Eq. (\ref{modSkellam}) is zero at all orders except zero, when it is unity. So the equation (\ref{delta}) and  condition (C5) are satisfied. 

To guaranty the symmetries (\ref{symM}) and above discussed limiting values, one has  to put $\alpha_1 = {\overline N}_2/(N_1+{\overline N}_2)$, ${\overline \alpha}_2 = N_1/(N_1+{\overline N}_2)$ and $\alpha_2={\overline N}_1/(N_2+{\overline N}_1)$, ${\overline \alpha}_1= N_2/(N_2+{\overline N}_1)$. Finally
 \begin{equation}
G_1={\overline G}_2=\frac{N_1{\overline N}_2}{N_1+{\overline N}_2},~G_2={\overline G}_1=\frac{N_2{\overline N}_1}{N_2+{\overline N}_1}
\label{G}
\end{equation}

Note, a common function  $Q(N_i, {\overline N}_i)$, which is symmetric under permutation $``1'' \leftrightarrow ``2''$ and vanish in the limits discussed in (C4),(C5), can be add to $G_i$, ${\overline G}_i$ without violation of all the properties discussed in (C1)-(C5) and values for $k_i^0$ (\ref{k0}). To fix it let us take into account the condition (C6). Then the simplest function $Q$ that guaranties  all the requirements is:
\begin{equation}
 Q= \left|k_1^0k_2^0\right|/2(N+{\overline N}).
\label{Q}
\end{equation}
 So, finally
\begin{equation}
M_i=G_i+Q, ~{\overline M}_i={\overline G}_i+Q.
\label{M}
\end{equation}

The modified Skellam distribution (\ref{modSkellam}) with the expressions (\ref{M}) for $M_i, {\overline M}_i$ and (\ref{k0}) for $k_i^0$ generalizes the original distribution (\ref{Skellam})  for semi-open (sub)systems ${N_i,\bar{N}_i}$
with the constraint for the total system $N-\bar{N}=B=const$. The  generalization is satisfied the obvious and necessary physical conditions (C1)-(C6). 

\begin{figure}[!ht]
   \centering
    \begin{minipage}[c]{.60\linewidth}
     \centering
     \caption{A comparision of the modified Skellam distri-\\bution (\ref{modSkellam}) with the Skellam-like one based on the \\Poissonian-binomial distribution (\ref{bin}), (\ref{F}), and also\\ with the original Skellam distribution.}
\label{fig:Fig1}
\end{minipage}%
\begin{minipage}[c]{.40\linewidth}
     \centering
      \includegraphics[width=\linewidth]{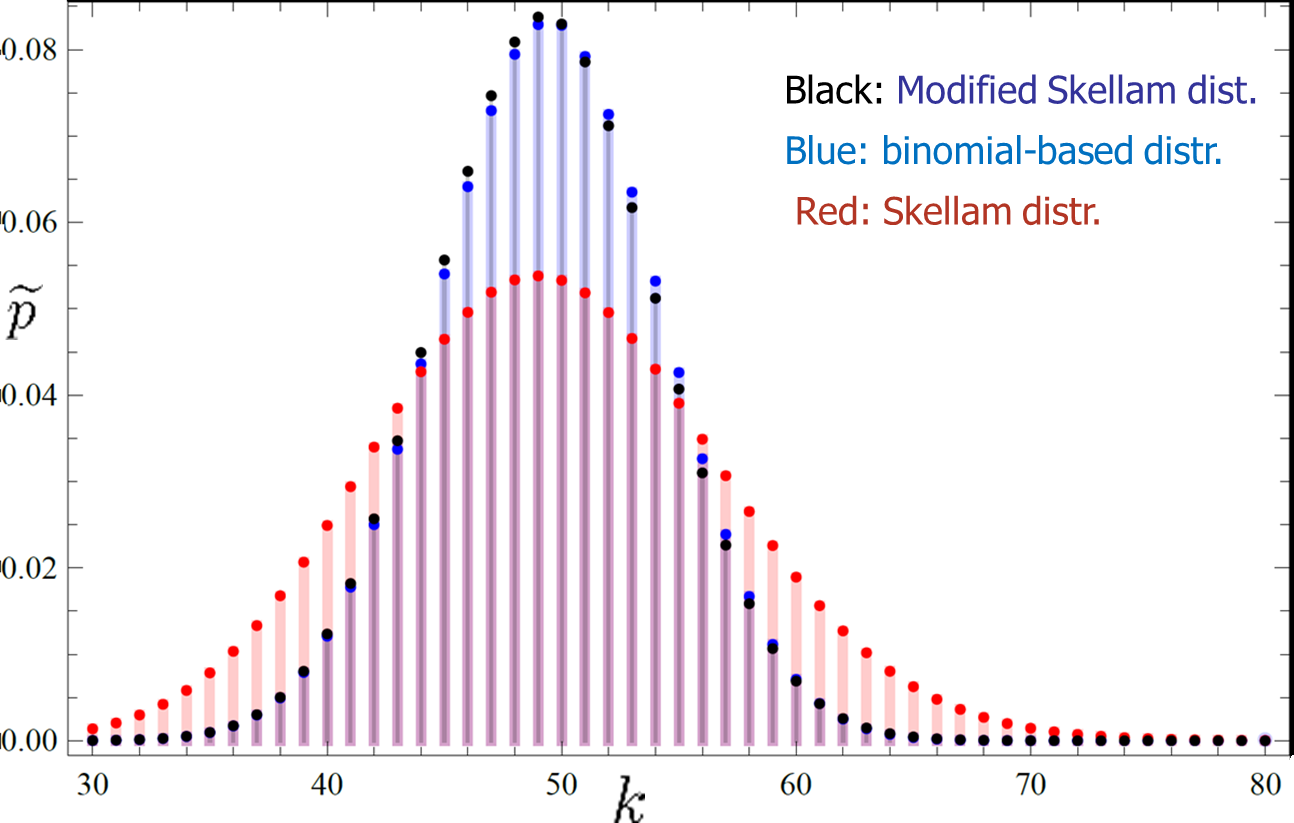}
\end{minipage}
\end{figure}

In Fig. \ref{fig:Fig1} we present, just for illustration, the comparison between modified Skellam distribution (\ref{Skellam}), Skellam-like binomial-based  distribution (\ref{bin}), (\ref{F}) and just Skellam distributions (\ref{Skellam}) at the same average values $N_i,{\overline N}_i$. The probabilities (\ref{q}) are $q=0.6$, ${\overline q}=0.5$, and $B=80$. One can see that the uncorrected for charge conservation law Skellam distribution is much wider than the ones accounting for this restriction.

Let us find for the semi-open $i$-subsystem the variance $\sigma^2$, skewness $S$ -- the measure of lack of symmetry of the probability distribution, and excess kurtosis $K$  -- the measure of the "tailedness". The calculations with modified Skellam distribution (\ref{modSkellam}), (\ref{k0}),  (\ref{M}) accounting for the corresponding mean values $m_i=N_i-{\overline N}_i$, see (C2), give the results   
\begin{equation}
\sigma_i^2=\sum_{k=-\infty}^{\infty} (k-m_i)^2 {\widetilde p}(k; N_i,{\overline N}_i) =(M_i+{\overline M}_i)\label{sigma}
\end{equation}

\begin{equation}
S_i=\sum_{k=-\infty}^{\infty}\left(\frac{k-m_i}{\sigma}\right)^3\widetilde{p}(k; N_i,{\overline N}_i)=\frac{M_i-{\overline M}_i}{\left(M_i+{\overline M}_i\right)^{3/2}}
\label{S}
\end{equation}

\begin{equation}
K_i=\sum_{k=-\infty}^{\infty}\left(\frac{k-m_i}{\sigma}\right)^4\widetilde{p}(k; N_i,{\overline N}_i)-3=\frac{1}{M_i+{\overline M}_i}
\label{K}
\end{equation}

Note, that the above equalities are exact, strictly speaking,  when $k_i^0$ are  integer numbers. Nevertheless, one can utilize the above expressions as the corresponding analytical continuations. Note, if $i$-system is a fairly small part of the total system (each of its  components), then $M_i\rightarrow N_i$, ${\overline M}_i\rightarrow {\overline N}_i$ and all the moments of the modified Skellam distribution are coincided with the known ones. If the subsystem, say ``2'', tends to the total system: $N_2\rightarrow N,~{\overline N}_2 \rightarrow {\overline N}$, then $M_1$ and  $M_2$ go to zero, so the dispersion $\sigma$ and all the other central moments, describing the fluctuations of the net baryon number, tends to zero. 

\begin{figure}[!ht]
   \centering
    \begin{minipage}[c]{.50\linewidth}
     \centering
     \caption{The ratios of the variance in the binomial-Poissonisan distribution (\ref{bin}), (\ref{F}) to the one (\ref{sigma}) in the modified Skellam distribution (\ref{modSkellam}). The red and black lines are related to the total net charge  $B=1$ and $B=150$ correspondingly at $q={\overline q}$; blue and green lines - to $B=150$ and $B=1$ correspondingly at $q={\overline q}+1/2$.}
\label{fig:Fig2}
\end{minipage}%
\begin{minipage}[c]{.50\linewidth}
     \centering
      \includegraphics[width=\linewidth]{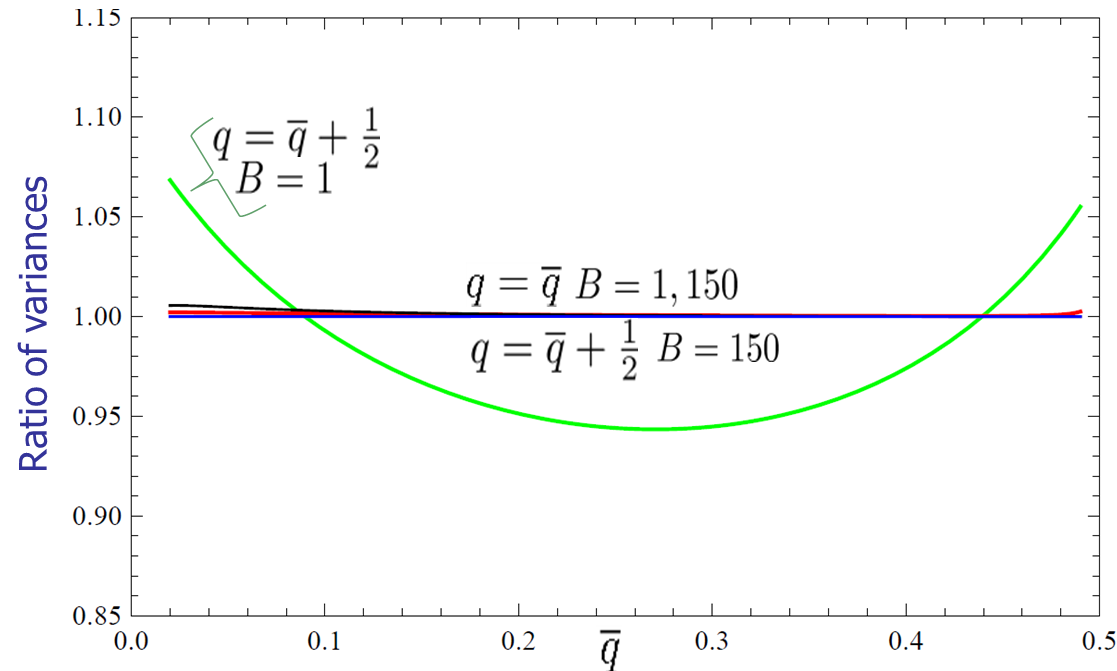}
\end{minipage}
\end{figure}

In the special case $q={\overline q}$ in (\ref{q}) the ratio of  $\sigma^2_i/\sigma^2_{i,Skellam}$  coincides (deviations not exceed 0.5\%)  with the result \cite{Anar} for this ratio,  $1-q$, obtained in the binomial-based model (\ref{bin}), (\ref{F}). In Fig. \ref{fig:Fig2} we demonstrate the ratios of variances $\sigma^2_{{\tt bin}}$, obtained from Eqs. (\ref{F}), (\ref{bin}) for the binomial-Poissonian distribution, and our result (\ref{sigma}) for $\sigma^2$, in the cases  when $q={\overline q}$ and also when $q={\overline q}+\frac{1}{2}$.
 Both situations are considered in the two limits: when the total net baryon numbers in the total system are $B=1$ and $B=150$. The maximal deviation between the models is reached 7$\%$ and achieved at small $B$ when $q={\overline q}+\frac{1}{2}$.    
 
\section{The modified Poisson and Gaussian distribution} 

\begin{figure}[!ht]
   \centering
    \begin{minipage}[c]{.80\linewidth}
     \centering
     \caption{The cartoon for the situation  when the only one component \\ in the selected subsystem ``1''  exists: ${\overline N}_1 = {\overline n}_1 =0$. Then one has \\ $k_1=n_1-{\overline n}_1=n_1\equiv n$, and the Poisson-like distribution ${\widetilde P}_{1+2}(n)$ at the \\total charge conservation law takes place.} 
\label{fig:ill4}
\end{minipage}%
\begin{minipage}[c]{.20\linewidth}
     \centering
      \includegraphics[width=\linewidth]{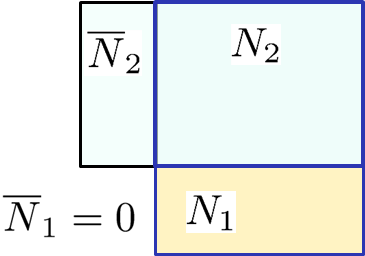}
\end{minipage}
\end{figure}

The analogy of the Poisson distribution ${\widetilde P}_{1+2}$ for  semi-open, say, ``$1$''-subsystem containing only single component, say, baryons $N_1$, is described by Eq. (\ref{modSkellam}) with ${\overline N}_1=0$. See Fig. \ref{fig:ill4}. So that ${\widetilde P}_{1+2}(n)={\widetilde p}(k_1=n_1-{\overline n}_1=n_1\equiv n; N_1,N_2,{\overline N}_2,{\overline N}_1=0)$.  

\begin{figure}[!ht]
   \centering
    \begin{minipage}[c]{.80\linewidth}
     \centering
     \caption{The cartoon for the situation  when in the selected subsystem \\ ``1''  the  component ${\overline N}_1 = {\overline n}_1 =0$, as well as the component $N_2 = n_2 =0$ \\ in the subsystem ``2'' . Then one has  $k_1=n_1-{\overline n}_1=n_1\equiv n$, and the \\modified Poisson distribution ${\widetilde P}(n)$ takes place.} 
\label{fig:ill5}
\end{minipage}%
\begin{minipage}[c]{.20\linewidth}
     \centering
      \includegraphics[width=\linewidth]{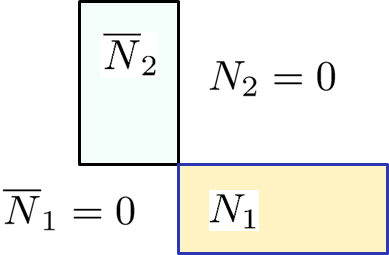}
\end{minipage}
\end{figure}

\begin{figure}[!ht]
   \centering
    \begin{minipage}[c]{.45\linewidth}
     \centering
     \caption{A comparision of the modified \\Poisson distribution (\ref{MP}) with the Pois-\\son-like one based on binomial-Poissonian \\ distribution (\ref{bin}), (\ref{F}), and also with the \\ standard Poisson distribution. The net \\baryon number $B=1$.}
\label{fig:Fig3}
\end{minipage}%
\begin{minipage}[c]{.50\linewidth}
     \centering
      \includegraphics[width=\linewidth]{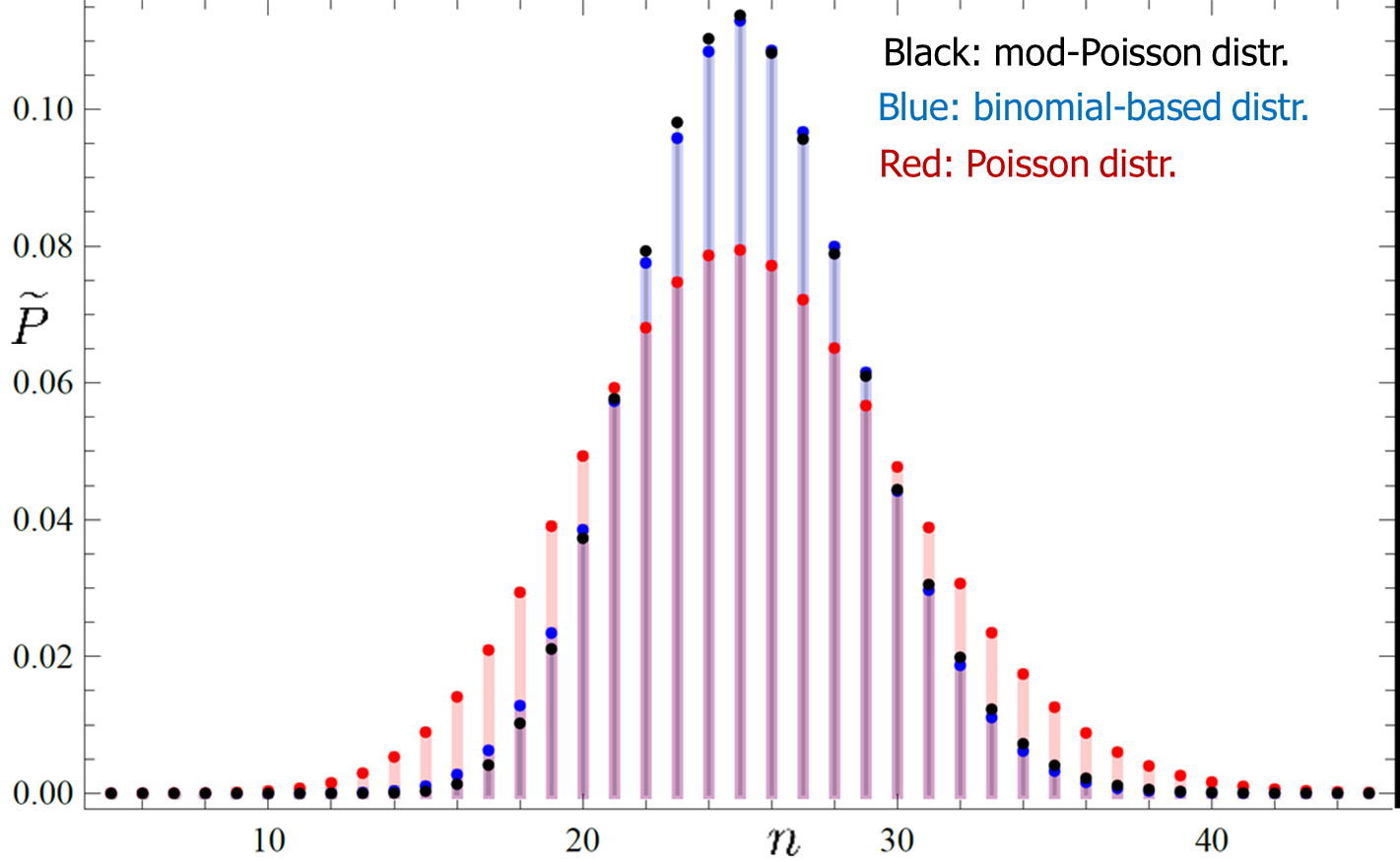}
\end{minipage}
\end{figure}

\begin{figure}[!ht]
   \centering
    \begin{minipage}[c]{.40\linewidth}
     \centering
     \caption{The ratios of variance of the binomial-based distribution (\ref{bin}), (\ref{F}) to the one obtained with the modified Poisson distribution (\ref{MP}) - blue points, and to the Poisson-like one ${\widetilde P}_{1+2}$ (see Fig. \ref{fig:ill4}) - red points, as the function of net baryon number $B$.}
\label{fig:Fig5}
\end{minipage}%
\begin{minipage}[c]{.55\linewidth}
     \centering
      \includegraphics[width=\linewidth]{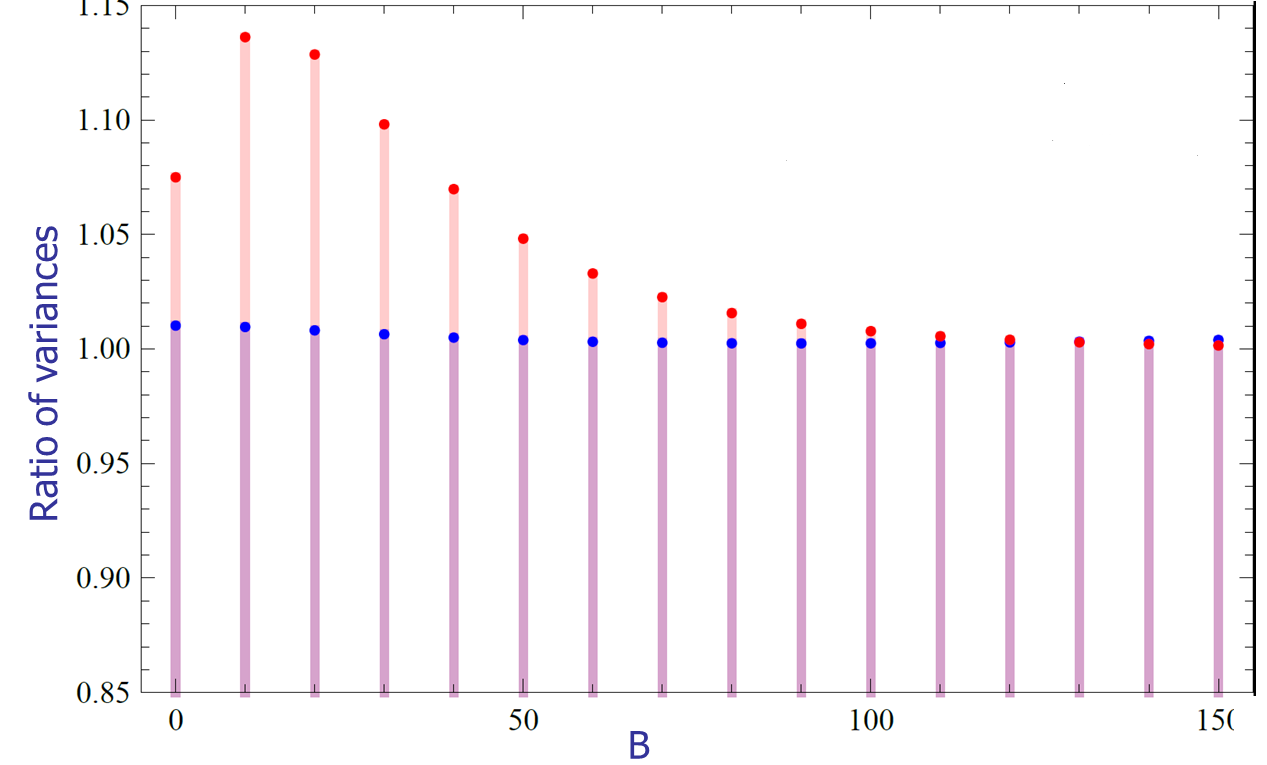}
\end{minipage}
\end{figure}

 Let us consider the case when the  subsystem ``2'' has also only one component, namely, ${\overline N}_2\neq 0$, so the only subsystems $N_1, {\overline N}_2\neq 0$ compose the total system, see Fig. \ref{fig:ill5}. Then the modified Poisson distribution is ${\widetilde P}(n)={\widetilde p}(k_1=n_1-{\overline n}_1=n_1\equiv n; N_1,{\overline N}_2,{\overline N}_1=0, N_2=0)$.
 A  significant analytical simplification in this case can be achieved if one neglects the term $Q$ (\ref{Q}) (it is, at least, one order of the value $(1/8)$ less than $M_1$) in Eq.(\ref{M}) for such a reduced system. Later we shall check such an approximation, see Figs. \ref{fig:Fig3}, \ref{fig:Fig5}. Then $M_i=G_i$. Denoting the mean  numbers of baryons $N_1 =N$ and antibaryons $\overline{N}_2 = \overline{N}$, one has $M_2\approx G_2=0$, $M_1\approx G_1 \equiv M = \frac{N\overline{N}}{N+\overline{N}}$, $k_1^0 \equiv k_0 = N - M$, and using the passage to the zeroth argument in Bessel function in Eq. (\ref{modSkellam}), one  can write the modified Poisson distribution, $\widetilde{P}$, accounting for the charge number conservation law in the system such as in Fig. \ref{fig:ill5}:
\begin{eqnarray}
\widetilde{P}(n;N)=\frac{e^{-M}M^{(n-k_0)}}{(n-k_0)!}, {\tt where}~  M = \frac{N\overline{N}}{N+\overline{N}},  k_0 = N - M. 
\label{MP}
\end{eqnarray}
As this was marked earlier, $k_0$ is a nearest integer number of the corresponding value. At $\alpha\equiv\frac{\overline{N}}{N+\overline{N}}\rightarrow 1$ the Eq. (\ref{MP}) coincides with the Poisson distribution (\ref{Poisson}), ${\widetilde P}(n;N)=P(n;N)$, at $\alpha \ll 1$ , ${\widetilde P}\rightarrow \delta^{N=B}_{n}$. In Fig. \ref{fig:Fig3} we demonstrate comparison of modified Poisson distribution  $\widetilde{P}(n;N)$ (\ref{MP}) with some other distributions at the same mean values $N, {\overline N}$ and $B=1$. One can see that the results  in our  simple approximation (\ref{MP}) and the binomial-Poissonian model (\ref{bin}), (\ref{F}) are fairly close, and at the same time the standard Poisson distribution (\ref{Poisson}) is much wider because does not take into account the charge conservation law.  
 
The modified Poisson distribution is normalized as easy to check.
The mean value $m=\left\langle n\right\rangle = N$. In the approximation (\ref{MP}) the variance $\sigma^2$, skewness $S$ and kurtosis $K$ are defined by the formulas (\ref{sigma})-(\ref{K}) with $M_1 = M$ and $M_2 =0 $. In Fig. \ref{fig:Fig5} we present the ratio of variances obtained in the modified Poisson function ${\widetilde P}$ (\ref{MP}) and in the corresponding binomial-Poissonian distribution (\ref{bin}) with probabilities  $q=1$ and ${\overline q}=0$, for a wide interval of net baryon charge $B$. We see a good agreement within 1\% between these two models. As for the Poisson-like distribution, ${\widetilde P}_{1+2}$, the deviation for the corresponding results can reach 14\% at relatively small $B$.

The transition to the correspondent Gaussian distribution is straightforward by means of the Stirling approximation and  standard procedure: $x=n=M(1+\delta); ~ N,M \gg 1,~ \delta \ll 1 $. Then one get from (\ref{MP}):
\begin{equation} 
\widetilde{P}(n;N)\rightarrow \frac{1}{\sqrt{2\pi M}}e^{\frac{(x-N)^2}{2M}}
\label{gauss}
\end{equation}
The limits are obvious: if $N \ll {\overline N}$, then $M\rightarrow N$ in (\ref{gauss}), if $N\gg {\overline N}$  ($N \rightarrow B$), then $\widetilde{P}(n;N)\rightarrow \delta (x-N)$. 

\section{Summary} 
A simple analytical generalization of the  Skellam distribution for an arbitrary two-component subsystem  accounting for  a charge-like conservation law in the total system is proposed. It is compared with the numerically evaluated binomial-Poissonian model, a very good agreement with previously found in such a model \footnote {for the particular case of equal probabilities to find baryon and anti-baryon in the selected sybsystem.} \cite{pbm}  the variation of the net baryon charge is observed. The results coinside within less than 0.5\%. The same concerns the case when the number of baryons is much more than anti-baryons, and wise-verse. Being consider in full region of baryon and antibaryon probabilities to belong the subsystem, the deviation in results of these two models do not exceed 15\%. The extremely simple approximations for the Poisson and the corresponding Gaussian distribution for considered type of systems are obtained based on the modified Skellam distribution. The presented formulas for the modified Poisson distribution are in a good agreement with numerical calculations in the binomial-Poissonian model, and so can be considered as a good analytic approximations for the later.  
  
	It is worthy noting that despite the closeness of the results, the analytic approach proposed in the note is fully independent and the simplest among the possible models  generalizing the Skellam and Poisson distributions for semi-open subsystems under the total charge-like conservation constraint.  The analytic expressions for variation, skewness and kurtosis generated by the  modified distributions are presented. The work is planning to apply for an analysis of different baryon \& anti-baryon observables in $pp$ and $AA$ collisions at high and intermediate energies.

\begin{acknowledgments}
Author is grateful to P. Braun-Munzinger and A. Rustamov for initializing discussions.   
The research was carried out within the scope of the EUREA: European Ultra Relativistic Energies
Agreement (European Research Network "Heavy ions at ultrarelativistic energies") and support by NASU, Agreement F-2018. 
The work is partially supported by NAS of Ukraine Targeted research program ``Fundamental research on high-energy physics and nuclear physics (international cooperation)''.
\end{acknowledgments}

\end{document}